\documentclass[conference]{IEEEtran}
\IEEEoverridecommandlockouts

\usepackage[backend=bibtex,style=ieee]{biblatex}
\usepackage{amsmath,amssymb,amsfonts}
\usepackage{algorithmic}
\usepackage{graphicx}
\usepackage{textcomp}
\usepackage{xcolor}
\usepackage{listings, listings-rust}
\usepackage{svg}
\usepackage{hyperref}

\newtheorem{definition}{Definition}

\def\BibTeX{{\rm B\kern-.05em{\sc i\kern-.025em b}\kern-.08em
    T\kern-.1667em\lower.7ex\hbox{E}\kern-.125emX}}
\begin{document}

\title{Finding a Crab in the C: Assured Translation via Comparative Symbolic Execution\\
\thanks{This material is based upon work supported by the Defense Advanced Research Projects Agency (DARPA) and the Naval Information Warfare Center (NIWC) Pacific, under Contract No. N66001-20-C-4018. The views, opinions and/or findings expressed are those of the author and should not be interpreted as representing the official views or policies of the Department of Defense or the U.S. Government. Distribution Statement ``A'' (Approved for Public Release, Distribution Unlimited).}
}

\author{\IEEEauthorblockN{Caleb Helbling}
\IEEEauthorblockA{\textit{Draper} \\
Cambridge, MA, USA \\
chelbling@draper.com}
\and
\IEEEauthorblockN{Graham Leach-Krouse}
\IEEEauthorblockA{\textit{Draper} \\
Cambridge, MA, USA \\
gleach-krouse@draper.com}
\and
\IEEEauthorblockN{Michael Crystal}
\IEEEauthorblockA{\textit{Draper} \\
Cambridge, MA, USA \\
mcrystal@draper.com}
}

\maketitle

\begin{abstract}
Modern high-assurance software systems development favors memory safe languages such as SPARK (ADA) or Rust. However, developers often encounter non-memory safe code (e.g., C) in legacy systems and libraries which would be prohibitively expensive or risky to re-write. In response, developers have begun turning to machine learning/AI systems and other automated code translators. Automated translation comes with its own risks, however. The original and ported code are not precisely the same, semantically - otherwise there would be no point in performing the translation. To reduce these risks, we have developed cozy, a comparative binary analysis tool that simultaneously analyzes a binary compiled from ``unsafe'' source code and a binary compiled from a translation of the source code to a memory safe language. cozy walks the developer through differences in the behavior of the two binaries, presenting each difference and asking the user to assess whether the difference is intentional (good) or erroneous. Outside of the flagged differences, the binaries are formally verified to be equivalent. Consequently, the review process guarantees equivalence modulo changes approved by the developer. cozy has applications to automated translation, bug correction, code reviews, operation authorization, and automatic translation.
\end{abstract}

\begin{IEEEkeywords}
Rust, C, verification, symbolic execution, comparison, differential, diffing
\end{IEEEkeywords}

\section{Introduction}
For many years, languages such as C and C++ have been the tool of choice for authors of low level or performance critical software. Unfortunately these languages suffer from memory safety issues, which have led to a wide range of security vulnerabilities. Rust has recently emerged as a compelling alternative for the development of low level applications by offering exceptional performance and control without compromising on memory safety. The push towards using memory safe languages has even been taken up by policy makers, with the National Security Agency (NSA), Cybersecurity and Infrastructure Security Agency (CISA), and the White House Office of the National Cyber Director (ONCD) all releasing reports urging adoption \cite{nsa-1, nsa-2,, cisa, oncd}.

Legacy codebases of C or C++ applications can be very large, and simply re-writing the entire codebase to Rust at once is often impractical. Instead, incremental re-writes are needed to slowly add memory safety. This translation can be done by some combination of manual translation, formal transpilation \cite{C2Rust}, or machine learning based translation \cite{safetrans, Weiss2025-pc, Luo2025-rm, Zhou2025-pl}. But even careful incremental rewrites are not necessarily meaning-preserving. Changes to program behavior can be introduced intentionally by fixing bugs (for example, memory safety violations) found in the original source, or unintentionally by erroneous translation.

Detecting these changes in program behavior is therefore growing in importance, and new automated methods are needed to obtain comprehensive analysis results. In this paper we will discuss our ongoing efforts to adapt our tool, cozy, for comparing C and Rust programs. Although originally designed for analysis of binary micropatches \cite{cozy}, cozy can be used to analyze generic binary software.

\section{Comparative Symbolic Execution}
\begin{figure*}
\includegraphics[width=\textwidth]{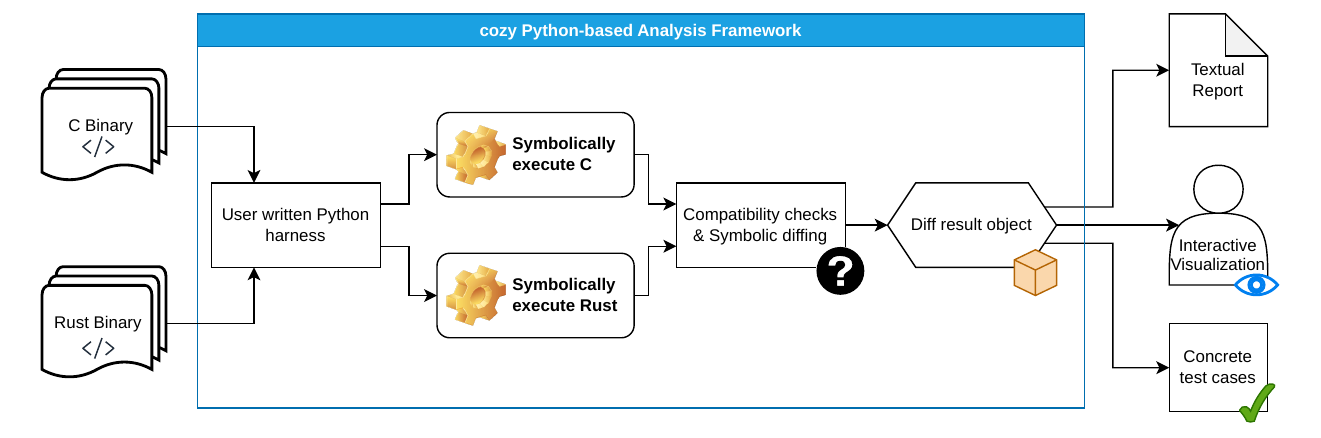}
\caption{The architecture of how cozy uses comparative symbolic execution in its analysis. The resulting diff object can be used to generate a textual report, exported for use in the interactive web based visualization software, or converted to concrete test cases for use in a testing suite.}
\end{figure*}

To make the comparison between two binary programs, cozy uses the angr symbolic execution framework \cite{angr} to perform \emph{comparative symbolic execution} \footnote{Prior work sometimes refers to ``differential symbolic execution'' \cite{differential_symbolic_execution} and ``relational symbolic execution'' \cite{farina}. Throughout this work, we use ``comparative symbolic execution'' as a generic term for this family of techniques.} (CSE) \cite{differential_symbolic_execution, farina}. During comparative symbolic execution, the pair of programs under analysis are fed the same symbolic input, and are then each symbolically executed to completion. During symbolic execution, the program state may ``split'' when control flow depends on the value of a symbolic variable, resulting in two separate child states each being generated as possible successors to a shared parent state. With repeated branching, an execution tree is created, with the initial execution state lying at the tree root, and terminal program states as tree leaves.

Typically at a branch point, one child will add a new constraint $C$, and the other child will add the negation $\lnot C$. As a result, during execution each state accumulates a list of constraints (interpreted conjunctively) which can be fed to an SMT solver such as Z3.  Solving the system of constraints associated with a given state produces a concrete input that would drive the program to enter that particular state during ordinary, non-symbolic execution.

After both programs are run to completion, cozy then proceeds in its analysis by comparing pairs of terminal states. We say that a terminal state $s$ from the first program and $s'$ from the second program are \textit{compatible} if there is a possible input that satisfies the joint constraints associated with $s$ and $s'$. Compatible states are ones that make sense to compare, since two compatible terminal states represent a possible pair of behaviors that would have been produced by the same input to the two programs under analysis.

\medskip

\begin{definition}[Compatibility]
  \label{def:compat}
  \small
  \[\textit{Compatible} \triangleq \{ (s, s') \mid \textit{compatible}(s,s') \}\]
  \begin{equation*}
    \begin{aligned}
    \textit{compatible}(s, s') \triangleq \texttt{is\_sat}(s.\texttt{constraints}\,\wedge\\
    s'.\texttt{constraints})
    \end{aligned}
  \end{equation*}
\end{definition}

\paragraph*{Unsat core optimization}
A na\"{i}ve way to compute the \textit{Compatible} set is to check all $n^2$ pairs of terminal states for joint satisfiability. cozy implements a memoization-based optimization to enhance performance. When \(s.\texttt{constraints} \wedge s'.\texttt{constraints}\) is unsatisfiable for a pair of states \((s,s')\), cozy computes the \textit{unsat core} and caches it. The unsat core is the minimal set of clauses for which the conjunction is unsatisfiable. Later, when we want to know if a new pair $(s, s')$ is compatible, we first check if any previously discovered unsat core is a subset of the joint constraints $s.\texttt{constraints} \land s'.\texttt{constraints}$. If this check succeeds, then the joint constraints are immediately unsatisfiable, and we can skip the expensive call to \texttt{is\_sat}. Since most state pairs are incompatible in practice, the unsat core optimization drastically reduces the number of SMT solver queries.

Once a pair of states is determined to be compatible by their constraints, we then begin the process of comparing the inner machine state of the pair. As an example, the memory contents $v$ and $v'$ at a certain address from the first and second program's terminal states respectively can be compared with a SMT query \texttt{is\_sat($s.\texttt{constraints} \land s'.\texttt{constraints} \land v \neq v'$)}.

A C vs Rust comparison requires flexibility, as the compiler for each language may produce code that differs in data layout and storage locations. To account for this difference, we introduce the concept of \textit{annotations}, which enables the assignment of names to specific memory or data locations. An annotation name is simply a sequence of strings and integers, typically derived from the names of struct fields and array locations. In the current implementation of cozy, annotations must be provided by the operator as part of the Python harness. By using annotations to map out semantic correspondences between memory and data locations, cozy is able to align comparisons across language boundaries.

\section{Visualization}
cozy exposes a GUI for interactive exploration of the symbolic execution trees. Direct manipulation of the symbolic execution results is important. Symbolic execution, even restricting attention to compatible paths, produces a lot of information. When analyzing symbolic execution results, the user needs a way to cut out extraneous noise. Typically, only a small subset of all of the possible paths through a program are of genuine interest. 

\subsection{Focusing Mechanisms}

The cozy GUI offers three mechanisms for focusing on the relevant parts of symbolic execution results: \textit{highlighting}, \textit{pruning}, and \textit{compression}.

Several types of program states that are likely to be significant are automatically highlighted in the GUI. These include states that raised errors during execution, states at which a syscall or SimProc (modeled function) call occurred, states at which the program exceeded user-specified boundaries on loop iteration, and states at which a user-provided assert or postcondition failed. Colors indicate different categories of potentially significant states.

Besides calling attention to relevant results, it can be helpful to filter out irrelevant results. Pruning works as follows: cozy prunes (hides) each branch unless it is ``interestingly related'' to a compatible branch in the facing tree, where the user specifies (via the GUI) which relationships are interesting. For example, the user can indicate that two branches are interestingly related when their terminal states have different memory contents; pruning will then leave only the branches that differ from at least one compatible branch of the facing tree in terms of their final memory contents. The relations that the GUI checks are symmetric, so if a branch $b$ survives pruning because it is partnered with a compatible branch $b'$, then $b'$ will survive as well. Therefore, pruning will never result in an orphaned branch.

 Finally, cozy allows compression: merging successive nodes that represent uninteresting or inevitable computation steps. There are two available compression levels: the user can (1) merge adjacent nodes that have identical constraints and (2) merge every node that has a unique child with that child node, eliminating all straight-line sequences of symbolic states. \footnote{Symbolic execution can add a constraint without branching when, for example, the result of adding the negation of the constraint is unsatisfiable.} Using compression visually reduces the size of the execution tree, easing the process of interactively exploring program behavior.
 
\begin{figure}[t]
\includegraphics[width=0.485\textwidth]{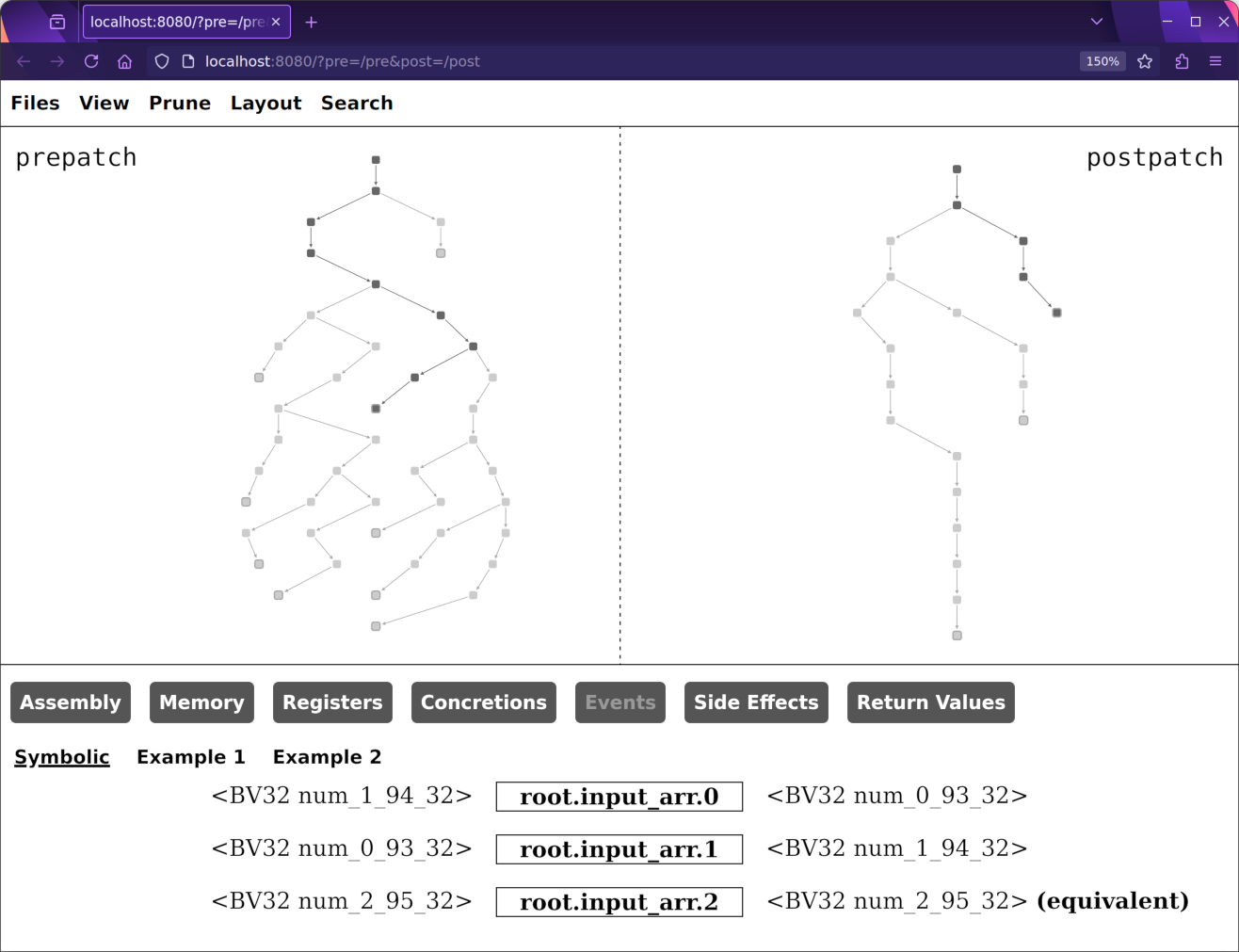}
\caption{In this example we interactively explore the execution trees of a C implementation of insertion sort with a Rust implementation of bubble sort containing a bug. After selecting one particular state pair from the execution tree, we observe in the diff panel that the mutated input arrays differ between the two versions (i.e., the symbolic \texttt{num\_0} and \texttt{num\_1} are placed in different locations in the mutated input array). This indicates a presence of a difference in program behavior.}
\label{fig:figure2}
\end{figure}

\subsection{The Diff Panel}

After finding the branches or state nodes of interest, cozy's user must be able to extract information from them. For examining individual nodes, we display features of symbolic state using simple tooltips. The more interesting challenge is displaying comparative information, when two compatible branches need to be examined simultaneously. For this, we use the \emph{Diff Panel}.

The sequence of nodes along a symbolic execution path corresponds to several different kinds of event streams: the stream of assembly instructions executed, the stream of read and write operations on memory and registers, and the stream of modeled IO effects. cozy compares these types of event streams using a familiar git-style line diff. For each type of event stream comparison, when the user mouses over an event in the diff panel, the UI highlights the tree node that corresponds to that event. This behavior enables the user to intuitively connect the contents of the tree-view to the contents of the event stream. 

In addition to diffing event streams, cozy supports comparisons of terminal states. For example, cozy can compare the final memory contents of two compatible branches. Figure \ref{fig:figure2} shows an example of using the memory tab of the diff panel to compare a pair of selected terminal states. If the symbolic expressions for the final memory contents of the states are equivalent, then cozy reports this fact; if they are not, cozy generates multiple \emph{concretions} that each illustrate a possible scenario in which the terminal states differ despite an identical initial state.

Finally, the diff panel can display concrete inputs that exercise each of the branches being compared. Compatibility guarantees the existence of at least one input that produces the two sequences of behaviors that the branches represent. The concretion view in the GUI's diff panel displays example inputs that are shared between the two compatible paths. This feature, in combination with cozy's pruning functionality, make it possible to recover specific inputs that generate execution paths of interest, especially paths where behavior differs interestingly between the two binaries being compared.

Compatibility does not guarantee that \textit{every} input that produces the behavior associated with the first branch also produces the behavior associated with the second branch, or vice versa. If we consider the set of concrete inputs that lead to a specific brach from the first program and the second program as $A$ and $B$ respectively, then the set difference $A \setminus B$ may be non-empty. In such scenarios, cozy reports inputs in the sets $A \setminus B$ and $B \setminus A$. Based on this information, we can determine if one branch ``refines'' the other, or if the input sets are exactly equivalent.

\section{Experimental Results}
To validate cozy's ability to compare C and Rust programs, we performed several experiments designed to gauge usability and performance characteristics. For each experiment, we designed a program and created a handwritten C implementation, a handwritten Rust implementation, and an implementation derived from the handwritten C using Galois/Immunant's automated C2Rust transpiler \cite{C2Rust}.

\begin{enumerate}
    \item In our first experiment, we implemented insertion sort, and were able to verify that the C, Rust, and C2Rust versions were equivalent. As a follow up to this test, we compared a C version of insertion sort with a Rust implementation of bubble sort. cozy was able to verify that the sorted outputs were equivalent, despite a difference in both implementation language and algorithm choice. Input array lengths were bounded by a small constant to ensure execution termination.
    \item In our second experiment, we wrote a simple watch program to test branching, arithmetic operations, and basic data structures. We first defined a \texttt{datetime} struct, which contained an hour, minute, second, month, day, year and day of week field. We then created a \texttt{second\_tick} function, which returned a new \texttt{datetime}, incremented by 1 second from the input. The bulk of the \texttt{second\_tick} function revolved around rollover of the different fields. Rollover occured in multiple places: seconds to minutes, minutes to hours, etc. Extra logic was required to account for the variable number of days in each month, and for the determination of the number of days in February for leap years.
    \item In our third experiment, we created a box blur algorithm, which returned a blurred version of an input image. A kernel size parameter was also used as input to control the size of box inside the algorithm (and therefore the blurriness of the resulting image). This experiment was created to exercise analysis of programs with nested loops and more complex data structures. We modeled the input image as an array of pixels from the \texttt{array2d} crate. This particular crate used a dynamically sized Rust vector, so modeling this more complex data structure was necessary to construct the input to feed to the program. Once again, we were able to verify that the box blur implementations were equivalent for small input images and kernel sizes.
\end{enumerate}

\section{Challenges}
In our second experiment we encountered some differences in program behavior, due to how arithmetic overflows are handled in C vs Rust. In the C programming language overflows are undefined behavior, whereas in Rust, the program will either panic (if compiled in debug mode) or perform two's complement overflow (if compiled in release mode). The net result is that the program outputs will almost always diverge due to differences in how edge cases are handled in arithmetic operations. Under the assumptions appropriate for the inputs to this kind of algorithm, these differences would not actually be exposed during execution; so, to improve our modeling of the watch program, we simply inserted some assumptions bounding the input (for example bounding the \texttt{years} field to sensible ranges), thereby constraining the program to never enter an overflow state. At this point, the Z3 solver was sufficiently constrained to prove the equality of the resulting program outputs.

On the software engineering side, we encountered challenges in adapting cozy to interact with the Rust ABI. To make accurate comparisons, the cozy Python library needs to know where to place data in preparation for program execution, and where the return results are placed once execution is complete. The angr library has some support for the C programming language, but offers no support for the Rust programming language. We ameliorated the lack of Rust support in three ways: (1) through judicious use of \texttt{repr(C)} annotations on data structures, (2) by requiring the use of \texttt{extern "C"} on functions to obtain a stable calling convention, and (3) by using DWARF debug data for memory layout information and to reduce the burden to provide type information. While we cannot claim that we support all Rust datatypes, we do have sufficient support for many basic use cases.

Besides challenges related to the ABI, we also encountered issues with the use of compiler optimizations. In our watch example, the Rust compiler was able to optimize one of the modulo calculations away, replacing it with an equivalent calculation using only bitwise and multiplication operations \cite{fast-division}. The GCC version did not use this particular optimization. When the Z3 solver was asked to verify the equivalence of the two implementations, it was unable to make progress due to the lack of understanding of the theory underlying this particular optimization. The situation was resolved by turning off all optimizations on all compilers, including MLIR optimization in \texttt{rustc}.

\section{Related Work}
One alternative for verifying program equivalence is to create a single harness function which runs the two functions in sequence, then assert equivalence in the return results. This can be accomplished in the Kani Rust bounded model checker \cite{kani} with a sequential equality harness, and some experimental features that enable bounded model checking of code linking to C:

\begin{lstlisting}[language=Rust, style=boxed]
fn sequential_equality_harness() {
    // Create a nondeterministic input
    let input_val: u8 = kani::any();
    let out_c = c_fun(input_val);
    let out_rust = rust_fun(input_val);
    assert!(out_c == out_rust);
}
\end{lstlisting}

A sequential equality harness approach was taken by Yang et al. in their VERT system \cite{vert} to verify equivalence of LLM translated C programs. Although this approach may look simple at first glance, it has some hidden downsides. Since the harness runs the two functions one after the other, a copy of the execution tree of the second function is made for every leaf of the execution tree of the first function. Even with early pruning of unsat paths, programs with \textit{degenerate compatibility} (i.e., all pairs are compatible) will result in unnecessary re-computation.

In addition to the issue with the blowup of the number of simulation paths, sequential equality harnesses do not give insights into all paths that lead to equality failure, only a fairly coarse success/failure signal, and perhaps a counterexample to equivalence. Because cozy has access to the details of symbolic state evolution, we can identify the branches that make sense to compare---the ones that accumulate compatible pairs of constraints---and directly inspect the course of each possible execution interactively or via textual report. cozy focuses not on \emph{whether} the functions are equivalent, but on \emph{how} they differ.

Bai and Palit took another approach with their work on RustAssure \cite{rustassure}. In RustAssure, the C and Rust programs were independently symbolically executed using KLEE, and the symbolic return values were dumped to an external file for analysis. After that point, no further queries to the SMT solver were made, and the return symbols were compared for similarity via a graph distance metric. Our approach differs as we use a compatible state check utilizing the path conditions and perform symbolic diffing using Z3. We believe that punting equality checking to the SMT solver is the most flexible approach and leads to more accurate results than a graph edit distance metric. Bai and Palit make a similar argument in their paper when assessing their approach, and note that moving to an SMT based approach is part of their future work.

\section{Discussion \& Conclusion}
In this paper we have described how our tool, cozy, can be used to compare C and Rust programs through symbolic execution. Comparison is achieved by compiling both programs to a common language---in this case assembly---and running the programs in angr. Our approach detects differences in all explored paths, and concrete examples can be extracted for further use in debugging or in a test suite. Since our framework is built as a Python library, it is easy to extract the reported data for downstream use. For example, a large language model could consume the textual report output from our tool to refine a C to Rust translation. Alternatively, a human operator can browse the program execution tree using our web based visualization software to quickly gain an understanding of program behavior.

Problems with execution performance are currently the most significant limitation in using cozy for comparing C and Rust. We should note here that the performance bottleneck is almost always in the interpretation of the binary program itself, not in the solving of constraints by the SMT solver. We suggest that the most fruitful path forward would be to port the strategies taken by cozy to a more performant symbolic execution engine for a suitable IR like Ghidra's P-code or Valgrind's VEX - perhaps an engine written in Rust rather than angr's Python.

Despite these scalability limitations, we are excited that our existing micropatch analysis tool could be so easily adopted to compare C and Rust programs. Our tool shows promise in not only cross-language comparison, but also cross-algorithm comparison. We believe that the techniques outlined in this paper could be extended with additional software engineering, leading to broader support of Rust features and capabilities.

\section{Mandatory Data Availability Statement}
The cozy comparative symbolic execution engine is available on our GitHub at \href{https://github.com/draperlaboratory/cozy}{draperlaboratory/cozy}, and as a PyPI package under the name \href{https://pypi.org/project/cozy-re/}{cozy-re}. The programs from the experimental results section are available on Zenodo at \href{https://zenodo.org/records/19669436}{10.5281/zenodo.19669436}.


\IEEEtriggeratref{11}
\printbibliography



\end{document}